\input{aipcheck}

\documentclass[final,nomathfonts]{aipproc}
\layoutstyle{6x9}

\usepackage[english]{babel}
\usepackage{amsmath}

\begin{document}
\global\long\def\f#1{\mathrm{f}^{#1}}
 \global\long\def\g{g_{\mathrm{c}}}
 \global\long\def\gf{\frac{\g^{2}}{f}}
 \global\long\def\te#1#2{T_{#1}\left(\hat{\mathbf{k}}_{#2}\right)}
 \global\long\def\mpl{m_{\mathrm{Pl}}}
 \global\long\def\e{\mathrm{e}}
 \global\long\def\jc{\varphi_{\mathrm{c}}}
 \global\long\def\gfi{\frac{\g^{2}}{f_{0}}}
 \global\long\def\fnl{f_{\mathrm{NL}}}

\title[$\zeta$ from non-Abelian Vector Fields]{Generating $\zeta$ with non-Abelian Vector Fields}

\author{Mindaugas Kar\v{c}iauskas}{address={CAFPE and Departamento de F\'isica Te\'orica y del Cosmos, Universidad de Granada, Granada-18071, Spain (mindaugas@ugr.es)}}

\begin{abstract}
In this paper the generation of the primordial curvature perturbation
by vector fields of general non-Abelian groups is discussed. We show
that non-Gaussianity of the perturbation is dominated by contributions
from superhorizon evolution of fields. Also we find that non-Abelian
vector fields of reasonably large groups can generate the total of
the curvature perturbation without violating observational constraints
on the angular modulation of the spectrum.
\end{abstract}
\maketitle
\section{Introduction}

Measuring statistical anisotropy alongside non-Gaussianity can give
very important information about the physics of the very early Universe.
It manifests itself as an angular modulation of the two-point and
higher order correlation functions, for example, the quadrupole modulation
of the spectrum $ $\citep{Ackerman_etal(2007)}\begin{equation}
\mathcal{P}_{\zeta}\left(\mathbf{k}\right)=\mathcal{P}_{\zeta}^{\mathrm{iso}}\left(k\right)\left[1+g_{\zeta}\left(\hat{\mathbf{n}}\cdot\hat{\mathbf{k}}\right)\right],\end{equation}
where $\hat{\mathbf{n}}$ is the preferred direction. Currently two
observational bounds exist in the literature on primordial $g_{\zeta}$:
$\left|g_{\zeta}\right|<0.3$ \citep{Groeneboom_etal(2009)anisotropy2}
and $\left|g_{\zeta}\right|<0.07$ \citep{Hanson_etal(2010)statAnis}.
The Planck satellite should better this bound with predicted accuracy
of $0.01$ ($2\sigma$) \citep{Ma_etal(2011)statAnis}. But even if
$g_{\zeta}$ is found to be negligible, the angular modulation
can still be dominant in higher order correlators \citep{Dimopoulos_us(2009)_fF2}.

Constraints on statistical anisotropy are especially useful in determining
possible effects of vector fields in the very early Universe. Their
non-negligible role could be expected as vector fields are ubiquitous
in particle physics models. Apart from other effects, they could have
generated the total or part of $\zeta$ \citep{Dimopoulos2006} or
influenced the dynamics of the inflaton \citep{Kanno_etal(2008)anisInfl}.
From a theoretical side, scenarios with vector fields open a new window
for the inflationary model building \citep{Dimopoulos_etal(2011)vFd-eta}.

In this work we show that non-Abelian vector fields of reasonably
large groups can also generate $\zeta$ without violating observational
bounds on $g_{\zeta}$. Moreover, its statistical properties are dominated
by perturbations of classical fields \citep{Karciauskas(2012)nAb1}.

\section{The Lagrangian}

We consider a Lagrangian of massless, non-Abelian vector fields with
time dependent kinetic function\begin{equation}
\mathcal{L}=-\frac{1}{4}f\left(t\right)F_{\mu\nu}^{a}F_{a}^{\mu\nu},\label{eq:Lagr-def}\end{equation}
where the field strength tensor $F_{\mu\nu}^{a}$ is given by\begin{equation}
F_{\mu\nu}^{a}=\partial_{\mu}A_{\nu}^{a}-\partial_{\nu}A_{\mu}^{a}+\g\f{abc}A_{\mu}^{b}A_{\nu}^{c}.\label{eq:Fmunu}\end{equation}
and $\mathrm{}$$\f{abc}$ are the Lie algebra structure constants
of any non-Abelian group. We assume that the gauge kinetic function
$f\left(t\right)$ is modulated by some of the dynamical degrees of
freedom of the full theory, which can be the inflaton itself \citep{Dimopoulos_etal(2011)vFd-eta}.
But in this work we do not need to specify the origin of this modulation.
Instead, we assume that $f\left(t\right)$ is not modulated by the
inflaton and that vector fields make a negligible contribution to
the total energy budget of the Universe.

Note, that vector fields $A_{\mu}^{a}$ in Eq.~\eqref{eq:Fmunu}
are defined with respect to the comoving coordinates \citep{Dimopoulos_etal_anisotropy(2008)}.
Spatial components of physical vector fields are given by $W_{i}^{a}=\sqrt{f}A_{i}^{a}/a$,
where we also canonically normalised the field. Also note, that the
self-coupling coefficient $\g/\sqrt{f\left(t\right)}$ of canonically
normalised fields is time-dependent.

In Ref.~\citep{Dimopoulos_us(2009)_fF2} it was shown that both degrees
of freedom of a massless Abelian vector field acquire a flat perturbation
spectrum $\mathcal{P}_{\delta W}=\left(H/2\pi\right)^{2}$ if $f\propto a^{-4}$.
In this case $f\gg1$ during inflation and self-couplings are suppressed,
which also results in flat perturbation spectra of $W_{\mu}^{a}$
fields.

\section{Statistically Anisotropic Perturbations}

Self-interactions of non-Abelian vector fields render their perturbations
non-Gaussian. We calculate the non-Gaussianity using the {}``in-in
formalism'' \citep{Maldacena(2002)}. Limiting ourselves only to
the tree level bispectrum it is enough to consider two contributions
to the unitary evolution of states. Writing the interaction Hamiltonian
as $\hat{H}_{\mathrm{int}}=\hat{H}_{\mathrm{int}}^{\left(3\right)}+\hat{H}_{\mathrm{int}}^{\left(4\right)}$
the second of these contribution is\begin{eqnarray}
\hat{H}_{\mathrm{int}}^{\left(4\right)} & = & a^{4}\left(\tau\right)\int\mathrm{d}^{3}\mathbf{x}\,\frac{1}{2}\gf\left(\f{abc}\f{ade}+\f{adc}\f{abe}\right)W_{i}^{b}\delta\hat{W}_{j}^{c}\delta\hat{W}_{i}^{d}\delta\hat{W}_{j}^{e},\label{eq:H4-def}\end{eqnarray}
where $\delta$ denotes the perturbation of the field. The first contribution
is proportional to the gradient and thus is suppressed on superhorizon
scales.

For the field operators we impose Bunch-Davies initial conditions.
Integrating the expectation value of the three point correlator we
find the three-point correlation function at the end of inflation
as \begin{equation}
g_{3}\left(\mathbf{k}_{1},\mathbf{k}_{2},\mathbf{k}_{3}\right)=-\left(2\pi\right)^{3}\delta\left(\mathbf{k}_{1}+\mathbf{k}_{2}+\mathbf{k}_{3}\right)\frac{2H^{6}}{\prod_{i}^{3}2k_{i}^{3}}\mathcal{T}_{lmn}^{fgh}\left(\hat{\mathbf{k}}_{1},\hat{\mathbf{k}}_{2},\hat{\mathbf{k}}_{3}\right)I\left(k_{1},k_{2},k_{3}\right),\label{eq:g3}\end{equation}
where group indices on $g_{3}$ are suppressed for brevity. First,
note that the three point correlator is anisotropic: \begin{eqnarray}
\mathcal{T}_{lmn}^{fgh}\left(\hat{\mathbf{k}}_{1},\hat{\mathbf{k}}_{2},\hat{\mathbf{k}}_{3}\right) & \equiv & W_{m}^{b}\te{lj}1\te{nj}3\left(\f{abh}\f{agf}+\f{agh}\f{abf}\right)+\nonumber \\
 &  & +W_{l}^{b}\te{mj}2\te{nj}3\left(\f{abg}\f{afh}+\f{afg}\f{abh}\right)+\nonumber \\
 &  & +W_{n}^{b}\te{lj}1\te{mj}2\left(\f{abg}\f{ahf}+\f{ahg}\f{abf}\right),\label{eq:T4-def}\end{eqnarray}
where tensors $\te{ij}{}\equiv\delta_{ij}-\hat{k}_{i}\hat{k}_{j}$
depend on the direction of wave vectors. The amplitude $I$ in Eq.~\eqref{eq:g3}
is given by\begin{eqnarray}
I & = & \gfi\frac{k_{\mathrm{t}}^{7}H^{-8}}{4!}\left[6\mathrm{e}^{4N_{k}}\left(\frac{1}{3}-K_{1}+K_{2}\right)+2\mathrm{e}^{2N_{k}}\left(K_{1}-3K_{2}-\frac{1}{5}\right)-\right.\nonumber \\
 &  & \left.-\left(\gamma+N_{k}\right)\left(\frac{1}{5}K_{1}-K_{2}-\frac{1}{35}\right)+\frac{1}{300}\left(625K_{2}-137K_{1}+\frac{1019}{49}\right)\right],\label{eq:I4-full}\end{eqnarray}
where $f_{0}$ is the initial value of $f$, $\gamma\approx0.577$
is the Euler-Macheroni constant and $k_{\mathrm{t}}\equiv k_{1}+k_{2}+k_{3}$,
$K_{1}\equiv\sum_{i>j}^{3}k_{i}k_{j}/k_{\mathrm{t}}^{2}$, $K_{2}\equiv\prod_{i}^{3}k_{i}/k_{\mathrm{t}}^{3}$.
The parameter $N_{k}\equiv-\ln\left(k_{\mathrm{t}}/a_{\mathrm{end}}H\right)$
is the number of e-foldings from the horizon crossing of $k_{\mathrm{t}}$
to the end of inflation. For cosmological scales $N_{k}\sim60$ and
the first term in the above expression is the dominant one, i.e. the
bispectrum is dominated by the classical contribution. Indeed, in
Ref.~\citep{Karciauskas(2012)nAb1} it was explicitly shown that
solutions of classical field equations give the same result as $g_{3}$
in Eq.~\eqref{eq:g3} with the dominant term in Eq.~\eqref{eq:I4-full}
given by \begin{equation}
I\approx\frac{g_{\mathrm{c}}^{2}}{f_{\e}}\frac{4\pi^{4}}{12H^{2}}\mathcal{P}_{\delta W}^{2},\end{equation}
where $f_{\mathrm{e}}$ is evaluated at the end of inflation.

We calculate the curvature perturbation $\zeta$ using the $\delta N$
formula and assume that only vector field contributions are considerable.
Since all vector fields satisfy the same equation of motion and structure
constants $\f{abc}$ are of the same order, it is reasonable to assume
that homogeneous values of all vector fields are of the same order
of magnitude and their orientations in space are random. If this is
the case, the power spectrum of $\zeta$ becomes\begin{equation}
\mathcal{P}_{\zeta}\left(\mathbf{k}\right)=\mathcal{N}\mathcal{P}_{\delta W}N_{W}^{2}\left[1-\frac{1}{\mathcal{N}}\sum_{a}^{\mathcal{N}}\left(\hat{\mathbf{W}}^{a}\cdot\hat{\mathbf{k}}\right)^{2}\right],\label{eq:Pz-gen}\end{equation}
where hats denote unit vectors, $N_{W}\approx\left|\partial N/\partial W_{i}^{a}\right|$
for all $a$ and $N$ is the number of e-foldings of unperturbed expansion.
$\mathcal{N}$ in the above is the number of vector fields. As we
can see, the angular modulation of $\mathcal{P}_{\zeta}$ is suppressed
by $\mathcal{N}$. Thus vector fields of a large enough non-Abelian
group can generate the curvature perturbation without violating observational
bounds on $g_{\zeta}$. In view of the two bounds on $g_{\zeta}$,
$\mathcal{N}\geq4$ or $15$ if $\left|g_{\zeta}\right|<0.29$ or
$0.07$ respectively.

\section{Generating Anisotropic $\zeta$ at the End of Inflation}

The end-of-inflation scenario with a vector field was first studied
in Ref.~\citep{Yokoyama_Soda(2008)}. Here we employ this scenario
in a toy example, motivated by particle physics models. Consider a
Lagrangian which is invariant under transformations of some non-Abelian
symmetry group $G$\begin{equation}
\mathcal{L}=\frac{1}{2}\partial_{\mu}\varphi\partial^{\mu}\varphi-\frac{1}{4}fF_{a}^{\mu\nu}F_{\mu\nu}^{a}+\frac{1}{2}\mathrm{Tr}\left[\left(D_{\mu}\Phi\right)^{\dagger}D^{\mu}\Phi\right]-V\left(\varphi,\Phi\right),\label{eq:GUT-L}\end{equation}
where $\mathrm{Tr}\left[\ldots\right]$ stands for trace, $\varphi$
is the inflaton field and $F_{\mu\nu}^{a}$ is the field strength
tensor defined in Eq.~\eqref{eq:Fmunu}. $\Phi$ in the above is
the Higgs field corresponding to a non-trivial representation of $G$
and the covariant derivative is $D_{\mu}=\partial_{\mu}+i\lambda_{A}\mathbf{T}^{a}A_{\mu}^{a}$,
where $\lambda_{A}$ is the gauge coupling constant and $\mathbf{T}^{a}$
are generators of the symmetry group $G$. Rewriting the Higgs field
as $\Phi\equiv\phi\mathbf{l}$, where $\mathbf{l}$ defines the direction
of symmetry breaking in the field space with $\mathrm{Tr}\left[\mathbf{l}^{\dagger}\mathbf{l}\right]=1$,
the potential $V\left(\varphi,\Phi\right)$ takes the form\begin{equation}
V\left(\varphi,\phi\right)=\frac{1}{4}\lambda\left(\phi^{2}-M^{2}\right)^{2}+\frac{1}{2}\kappa^{2}\varphi^{2}\phi^{2}+V\left(\varphi\right).\label{eq:V-hybrid}\end{equation}
One can easily recognize this as the potential of hybrid inflation
in which $\phi$ is stabilised at the origin during inflation. But
in contrast to the standard hybrid inflation, in this example $\zeta$
is generated by gauge fields at the end of inflation. To see this,
note from Eqs.~\eqref{eq:GUT-L} and \eqref{eq:V-hybrid} that the
effective mass squared of the Higgs field $\phi$ is

\begin{equation}
m_{\mathrm{eff}}^{2}\left(\mathbf{x}\right)=\kappa^{2}\varphi^{2}-\lambda M^{2}-\lambda_{A}^{2}A_{\mu}^{a}A_{b}^{\mu}\mathbf{l}^{\dagger}\mathbf{T}^{a}\mathbf{T}^{b}\mathbf{l}.\label{eq:m-eff}\end{equation}

In this scenario the curvature perturbation power spectrum in Eq.~\eqref{eq:Pz-gen}
becomes\begin{equation}
\mathcal{P}_{\zeta}\left(\hat{\mathbf{k}}\right)\approx\lambda_{A}^{4}\mathcal{N}\mathcal{P}_{\delta W}\left(CW\right)^{2}\left[1-\frac{1}{\mathcal{N}}\sum_{\bar{a}}\left(\hat{\mathbf{W}}^{\bar{a}}\cdot\hat{\mathbf{k}}\right)^{2}\right],\label{eq:Pz-end-of-infl}\end{equation}
where we used $W\approx\left|\mathbf{W}^{\bar{a}}\right|$ for all
$\bar{a}$ and indices with overbar run only over generators broken
at the end of inflation. Also in the above $C\equiv N_{\e}/\kappa^{2}f_{\e}\jc$,
where $\varphi_{\mathrm{c}}$ is the critical value of the inflaton
and $N_{\e}\equiv\partial N/\partial\jc$. Note that $\mathcal{N}$
in Eq.~\eqref{eq:Pz-end-of-infl} is the number of \emph{massive}
vector fields.

As was noted before, the bound on $\mathcal{N}$ is $\mathcal{N}\geq4$
or $15$. Assuming $G$ to be a special unitary group $SU\left(N\right)$
which at the phase transition is broken to $SU\left(N-1\right)$,
the symmetry breaking results in $\mathcal{N}=2N-1$ massive gauge
fields. Thus, a weaker bound on $\mathcal{N}$ is satisfied with the
$SU\left(3\right)$ group while the stronger bound requires $SU\left(8\right)$.
Therefore $\zeta$ can be generated by gauge fields of reasonably
large groups.
\begin{description}
\item [{Acknowledgments}] MK is supported by \textbf{CPAN} CSD2007-00042
and \textbf{MICINN} (FIS2010-17395) grants.
\end{description}
\bibliographystyle{aipprocl}
\bibliography{Ref_ERE2011Procs}

\end{document}